\def \etal   {{et al.\thinspace}}
\def\gtorder	{\mathrel{\raise.3ex\hbox{$>$}\mkern-14mu\lower0.6ex\hbox{$\sim$}}}
\def\ltorder	{\mathrel{\raise.3ex\hbox{$<$}\mkern-14mu\lower0.6ex\hbox{$\sim$}}}

\documentstyle[11pt,paspconf,epsf]{article}

\begin{document}

\title{The Shape of Dark Matter Halos}

\author{Penny D. Sackett}
\affil{Kapteyn Astronomical Institute, 
9700 AV Groningen, The Netherlands}

\begin{abstract}

Techniques for inferring the radial and geometric form of dark matter 
halos and the results they have produced to date are reviewed. 
Dark halos appear to extend to at least $\sim$50~kpc with   
total enclosed masses that rise linearly with radius R.  Whether this behavior 
can be extrapolated to distances as large as 200~kpc and beyond 
is controversial; results at this radius are model-dependent.  
Observationally, the geometrical form 
of the dark halo can be characterized by the equatorial axis ratio 
$(b/a)_\rho$ (ovalness) and vertical-to-equatorial  
axis ratio $(c/a)_\rho$ (flattening) of the total density. 
Different techniques consistently yield $(b/a)_\rho > 0.7$ 
(and thus $(b/a)_\Phi> 0.9$) at R$\sim$20~kpc, with more axisymmetric 
values, $(b/a)_\rho \gtorder 0.8$, being more likely.  
Results are less consistent for the vertical flattening, perhaps 
due to the difference in the spatial regions probed by different techniques 
or inappropriate assumptions.  Techniques that probe furthest from the 
stellar plane ($z\sim$15~kpc) consistently implicate substantially flattened 
$(c/a)_\rho = 0.5 \pm 0.2$ dark halos.  
These axis ratios are in acceptable agreement with expectations 
from N-body simulations of cold 
dark matter mixed with $\sim$10\% dissipational gas. 
\end{abstract}


\section{Introduction}

A large portion of galactic dynamics is the study of the 
relationship between mass and motion in and around galaxies.  
The total gravitational potential of galaxies 
--- its extent, radial density profile, flattening, and trixiality ---
strongly affects the motions and morphology of associated  
stellar and gaseous components, and thus subjects as varied as 
merging rates, the evolution of galactic disks, bars, warps and 
spiral structure, the Tully-Fisher and Fundamental Plane relations, 
and the nature of the presumed galactic dark matter constituents.   

To the extent that the strength of the gravitational potential of galaxies 
is provided by unseen dark matter --- rather than by poorly characterized 
luminous mass or a modification to the Law of Gravitation 
(McGaugh 1999, and references therein) --- 
the morphology and motion of luminous galactic tracers 
can be used to deduce the form of the underlying dark mass 
distribution, provided that the relationship linking them is 
sufficiently well understood. 
Each luminous tracer has its own advantages and disadvantages for this 
purpose; many are suitable only for galaxies of a given type or within a given
galactocentric radius. 

Several observational techniques and the results that they have produced 
regarding dark halo shape are reviewed here, with emphasis on the 
strengths and weakness of each.  Many of the results are model-dependent; 
some are in apparent conflict with others.   In this brief and necessarily 
incomplete review, it will not be possible to do justice to the complexity 
and richness of the scientific dialogue on these issues or to resolve 
apparent discrepancies. Instead, emphasis will be placed on the assumptions 
and possible systematic effects inherent to various techniques, and on 
the areas of general agreement and controversy between them.  
The focus will be on spirals rather than early type galaxies, in part 
because the shape of dark halos is (generally) better constrained 
in disk galaxies but also because the situation with respect to 
ellipticals is reviewed elsewhere in this proceedings (Bridges 1999).  
For other reviews on the shape of dark halos, see Rix (1996) and 
Sackett (1996). 

\section{What do we mean by halo shape?} 

The shape of the dark halo is determined by its 
radial profile and --- to the extent that it can be described as 
triaxial --- the axis ratios $b/a$ (intermediate-to-long) and $c/a$ 
(short-to-long).  As we shall see, most all observational 
studies indicate that, in the region where they can be probed,   
dark halos are only mildly triaxial and oblate ($b \approx a$), 
with the equatorial plane of the dark halo nearly coinciding 
with that of the stellar body. 
Throughout this review, we will characterize the geometric form 
of the dark halo by $(b/a)_\rho$ and $(c/a)_\rho$, the  
axis ratios of {\it density\/}, not potential.
The gravitational potential $\Phi$ to which 
any density $\rho$ gives rise will in general be more spherical; 
a rough rule of thumb is
\begin{equation}
1 - (c/a)_\rho \equiv \epsilon_\rho \approx 3 \, \epsilon_\Phi 
	\equiv 3 \left[ 1 - (c/a)_\Phi \right]~~.
\end{equation}

\noindent
To first approximation, closed loop orbits of a tracer 
population will have the same ellipticity as the underlying potential, 
but their long axis will 
be perpendicular to that of $\Phi$. 

In the most general case, $(b/a)_\rho$ and $(c/a)_\rho$, which 
describe the ovalness and flattening of the dark density respectively, 
may be functions of radius.  
At a given radius, the shape of the total gravitational potential may 
differ from that due to the dark halo alone, depending on the relative 
shape and strength of the luminous and dark contributions to the total 
$\Phi$ at that position.

\section{Radial Profile of the Dark Halo}

The mass of the halo within a given enclosed radius 
depends strongly on the form of its radial density profile $\rho(r)$.  
The inner radial profile of the dark halo affects the 
dynamic interplay between dark matter and mature stellar systems, 
while the outer profile influences the dynamics of the 
galaxy-satellite system and its interaction with its 
nearest neighbors.

\subsection{How much dark mass is in the central regions of galaxies?} 

The quality of galactic rotation curves allows the accurate 
determination of the total (luminous + dark) 
mass enclosed within a radius equal to one-half the maximum 
extent $R_{\rm TRACER}$ of the velocity tracer (Sackett 1997).  
If the isodensity contours of total mass are not 
concentric ellipsoids, as they are unlikely to be when composed 
of several components of differing shape, uncertainties in 
the extrapolation of the outer rotation curve translate into uncertainties 
in the inferred enclosed mass beyond $R_{\rm TRACER}$. 
Interior to $R_{\rm TRACER}$, uncertainty in the {\it dark\/} density profile 
$\rho(r)$ is not due to uncertainty in dark halo 
geometry or inadequate velocity data, but to the uncertain 
mass-to-light ratio $M/L$ of the luminous component and thus the  
contribution of the luminous mass to the inner rotation support 
of the galaxy.  

The ``maximum disk hypothesis,'' which assumes 
that the mass of the stellar disk alone (possibly in conjunction 
with a stellar bulge) is responsible for the inner rotation speed of 
spirals, was first formulated in order to place a conservative 
lower limit on the amount of dark matter deduced from rotation 
curves (van~Albada \& Sancisi 1986).  Since then, evidence both 
for and against the suitability of the hypothesis as a description 
of the actual distribution of stellar mass in the interiors 
of bright, high surface-brightness spirals has been put forward 
(e.g., Palunas \& Williams 1999; Courteau \& Rix 1999, Sellwood 1999, 
and references therein).  
Mounting evidence suggests that dwarfs and low surface brightness spirals 
are dominated by dark matter at all radii (de~Blok \& McGaugh 1997), 
though this conclusion may not be universal (Swaters 1999).

\subsection{How far do dark halos extend?}

In spirals, the neutral hydrogen gas layer has an outer radius of 
30~kpc; planetary nebulae and globular clusters can be used to probe similar 
radii in ellipticals (Bridges 1999).  These tracers are generally 
consistent with an isothermal {\it total} 
mass distribution ($M(<R) \propto R$).  
In order to probe the outer radial profile and total extent of dark 
matter halos, a visible tracer at larger radii is required (Table~1).

Model potentials of the Milky Way built to support the 
orbits inferred from radial velocities and proper motions of the 
Magellanic Clouds and Stream yield a total of the Milky Way of 
$(5.5 \pm 1) \times 10^{11} {\rm M}_\odot$ inside 100~kpc 
(Lin, Jones \& Kremola 1995), about half of which 
must lie outside the present Cloud distance ($\sim$50~kpc).  
To reconcile this with larger estimate of 
$(5.4 \pm 1.3) \times 10^{11} {\rm M}_\odot$ inside 50~kpc 
necessitated by the enforcement of Local Group timing constraints 
and inclusion of Leo I as a satellite (Kochanek 1996), requires that the 
Galaxy have a dark mass of $\sim2.5 \times 10^{11} {\rm M}_\odot$ 
and a total of mass of $\sim3.0 - 3.5 \times 10^{11} {\rm M}_\odot$ 
interior to 50~kpc (Sackett 1996).  
Such a mass distribution would have a lower circular velocity 
--- as advocated by many authors, most recently Merrifield \& Olling (1998) --- 
than the current IAU standard value of 220~km~s$^{-1}$. 

The dark distribution of the Galaxy is thus 
likely to extend to at least 100~kpc.  Satellite kinematics  
have been used to argue that the mass distribution 
of both the Milky Way (Zaritsky 1998) and external galaxies 
(Zaritsky \& White 1994) extends isothermally 
to more than 200~kpc, though this conclusion is 
in conflict with that of an earlier, smaller study that combined 
information from 
rotation curves and satellites to conclude that galactic potentials 
were already keplerian at 3 disk radii (Erickson, Gottesman \& Hunter 1987).

\begin{table}
\caption{Inferred Extent of Dark Halos\tablenotemark{a}}\label{extent.table}
\begin{center}\scriptsize
\vglue -0.4cm\begin{tabular}{lcrrr}
 & & \multicolumn{3}{c}{Inferred Quantities} \\
\multicolumn{1}{l}{Method/Tracer} & \multicolumn{1}{c}{System}  
& \multicolumn{1}{c}{Extent} & \multicolumn{1}{c}{Mass} &  
\multicolumn{1}{c}{$M/L$\tablenotemark{b}} \\
~~~~Reference & & \multicolumn{1}{c}{(kpc)} 
& \multicolumn{1}{c}{($10^{11} M_\odot$)} & 
\multicolumn{1}{c}{($M_\odot / L_\odot$)} \\
\tableline
 & & & & \\
Magellanic Cloud and Stream Orbits & & & & \\
~~~~Lin, Jones \& Kremola 1995 & Milky Way 
                & $\gtorder 100$ & 2.8 ($<$~50\,kpc) & ---\\
    &           &                & 5.5 ($<$100\,kpc) & --- \\
Satellite Kinematics & & & & \\
~~~~Kochanek 1995 & Milky Way & $\gtorder 100$ & 5.4 ($<$~50\,kpc) & ---\\
                  &           &                & 8.0 ($<$100\,kpc) & --- \\
~~~~Erickson, Gottesman \& Hunter 1987 & Nearby spirals 
 & $\ltorder 100$ & --- & 20 \\
~~~~Zaritsky \& White 1994 & Distant spirals 
 & $\gtorder 200$ & $\sim$20 ($<$200\,kpc)  & $110 - 340$ \\
Giant Ring Kinematics & & & & \\
~~~~Schneider 1995 & Leo HI Ring 
 & $\ltorder 60$ & 5.6 ($<$100\,kpc) & 25\\
Tidal Tail Length & & & & \\
~~~~Dubinski, Hernquist \& Mihos 1996 & Simulations 
 & $\ltorder 100$ & --- & --- \\
Statistical Weak Lensing & & & & \\
~~~~Brainerd, Blandford \& Smail 1996 & Lum-Selected 
 & $> \, 30$ & 13$^{+16}_{-7}$ ($<$130\,kpc) & --- \\
\end{tabular}
\end{center}
\vglue -0.8cm
\tablenotetext{a}{H$_0 = 75$~km~s$^{-1}$~Mpc$^{-1}$ used throughout.}
\tablenotetext{b}{Mass-to-Light ratio within maximum extent.}
\end{table}

Satellites at the largest distances (200-300~kpc) from their 
primaries, where they are sure to probe the dark potential, have 
orbital times are on the order of a Hubble time; their non-equilibrium 
kinematics and distribution can only be studied in the context of 
a halo formation model.  The advantage of a cold, smooth tracer like 
HI gas over a hot discrete collection of satellites is that the 
orbit structure can be traced on the sky adding to the information gained 
from radial velocities.   The HI ring in Leo completely encircles 
the early-type galaxies M105 and NGC~3384 at a radius of 100~kpc.  
The radial velocities and spatial distribution 
of the gas are consistent with a single, elliptical {\it keplerian\/} 
orbit with a center-of-mass velocity equal to the centroid of 
the galaxy pair, and a focus that can be placed at the barycenter of 
the system without compromising the fit (Schneider 1995), 
suggesting that the dark matter in this system does not extend much 
beyond the ring pericenter radius of 60~kpc.  

The presence of long ($50 - 100$~kpc) tidal tails seen in 
HI emission in many interacting systems has been used to argue that dark 
halos must be more compact and/or less massive than those predicted 
by standard ($\Omega = 1$) cold dark matter cosmologies or inferred 
from the basis of satellite studies (Dubinski, Mihos \& Hernquist 1996;  
Mihos, Dubinski \& Hernquist 1998).  In particular, it was  
claimed that the dark halo could contribute no more than $\sim$90\% of 
the total enclosed mass.  This has been challenged by Springel \& White 
(1998) who argue that if the size and angular momentum of the disk 
and halo are linked, massive halos may generate long tidal tails because 
the disks are larger as well.  Dubinski, Mihos \& Hernquist (1999) 
agree with this conclusion, but conclude that the falling rotation curves 
then required at the edges of disks to produce long tidal tails may 
be at odds with large isothermal halos inferred by other methods.

The tangential distortion of background 
galaxies due to weak lensing by foreground galaxies is statistically 
measurable for large samples of source-lens pairs and high quality 
data.  Brainerd, Blandford \& Smail (1996) obtain 
a mass of 13$^{+16}_{-7} \times 10^{12}$~{\rm M}$_\odot$ 
within 130\,kpc with this method, although the 1$\sigma$ uncertainty 
on the truncation radius allows halo extents as small as 30~kpc.

\section{Non-axisymmetry of Dark Halos: $(b/a)_{\large \rho}$}

Are spiral galaxies axisymmetric ($b$ = $a$) or are the 
equidensity contours oval in the stellar equatorial plane? 
Table~\ref{bona.table} lists observational attempts to answer 
this question, where $(b/a)_\rho$ is the inferred axis ratio of 
the isodensity contours at the indicated radius.

Assuming that intrinsic ovalness of the stellar disk can be used to 
trace the non-axisymmetry of the underlying density, 
the statistical distribution of observed axis ratios on the sky 
can be related to $(b/a)_\rho$: the more elongated the density, 
the fewer galaxies will be observed with $(b/a)_{\rm LUM} \approx 1$.  
Recent attempts to use this method 
(e.g., Lambas, Maddox \& Loveday 1992; Fasano \etal\ 1993) 
with the APM and RC3 surveys find that while 
the stellar bodies of ellipticals and early type spirals 
appear to have some degree 
of triaxiality, spirals are more axisymmetric.  
A mean deprojected axis ratio of $(b/a)_{\rm LUM} = 0.9$ was found for  
isophototal contours of the APM spirals, 
and a sample of unbarred RC3 spirals yielded $(b/a)_{\rm LUM} = 0.93$.  
Assuming that the stellar distribution traces the potential of 
the underlying mass distribution (Eq.~1), and mindful of the 
complications of bar and spiral structure, the lower 
limit on the axis ratio of dark isodensity contours in 
the plane can be taken as $(b/a)_\rho \gtorder 0.7 - 0.8$.  

If perfectly face-on systems could be found and their axis ratios 
measured in a wavelength that is free of contamination from dust and 
the spiral structure caused by bright young populations, 
a more reliable estimation of the 
axisymmetry the potential might be obtained.  Rix \& Zaritsky (1995) attempted 
this using K band imaging and small HI linewidths as a selection for 
face-on systems.  The result indicates a small but persistent 
non-axisymmetry of $(b/a)_\Phi = 0.955$ in the potential corresponding to 
$(b/a)_\rho \approx 0.85$ for the underlying density distribution.

\begin{table}
\caption{Observational Constraints on Dark Halo 
Non-axisymmetry\tablenotemark{a}}
\label{bona.table} 
\begin{center}\scriptsize
\vglue -0.4cm\begin{tabular}{lcrr}
\multicolumn{1}{l}{Method/Tracer} & \multicolumn{1}{c}{System}  
& \multicolumn{1}{c}{R Extent Probed} & \multicolumn{1}{c}{($b/a$)$_\rho$} \\
~~~~Reference & & \multicolumn{1}{c}{(kpc)} & \\
\tableline
 & & & \\
Sky distribution of axis ratios & & & \\
~~~~Lambas, Maddox \& Loveday 1992 & APM Survey Spirals & Opt Radius & $\gtorder$0.7 \\
~~~~Fasano \etal\ 1993 & Unbarred RC3 Spirals & Opt Radius & $\gtorder$0.8 \\
K Band Imaging & & & \\
~~~~Rix \& Zaritsky 1995 & Small HI Linewidths & K Band Radius & 0.77-0.93 \\
Gas and Star Kinematics & & & \\
~~~~Kuijken \& Tremaine 1994 & Milky Way & 8 - 16~kpc & 0.75 \\
HI Ring Kinematics & & & \\
~~~~Franx, van~Gorkom \& de~Zeeuw 1994 & E/S0 IC 2006 & $\sim$13~kpc & $0.96 \pm 0.08$ \\
Harmonic Analysis of Gas Kinematics & & & \\
~~~~Schoenmakers 1998, 1999 & HI-rich Spirals & $1 - 2$ Opt Radii & $0.85 \pm 0.04$ \\
Low Scatter in Tully-Fisher Relation & & & \\
~~~~Franx \& de~Zeeuw 1992 & Spirals & HI Gas Radius & $\gtorder$0.84 \\
\end{tabular}
\end{center}
\tablenotetext{a}{For some references, $(b/a)_\rho$ has been estimated from 
$\epsilon_\Phi$ using Eq.~1.}
\vglue -0.7cm
\end{table}

The neutral hydrogen layer in spirals extends to comparable or 
larger radius than the light and thus might be expected to be a better 
probe of the axisymmetry of the dark halo.  At these radii, the 
gas is further from the disturbances in the potential caused by 
bars and stellar spiral structure (though spiral structure is often 
seen in the HI gas itself).  The cold dynamical state of 
HI and its dissipational nature makes it more likely to follow 
the closed loop orbits 
that are a good indication of isopotential contours in the disk plane.

Kuijken \& Tremaine (1994) have suggested that part of the 
apparent discrepancy between the rotation curves constructed from 
stellar and gaseous tracers in the Milky Way may be due to the fact 
that they are located at different positions within an inherently 
non-axisymmetric potential.  Their attempts to resolve these 
discrepancies requires that the line-of-sight to the Sun lie near 
the short axis of the potential and leads to an estimate of 
$\epsilon_\Phi = 0.08$ for 
the equipotential ellipticity at $ 1 -2 \, R_0$ (8-16~kpc), corresponding 
to $(b/a)_\rho = 0.75$.

Assuming that the HI gas ring in the E/S0 IC~2006 lies in the 
galaxy's symmetry plane, its shape and kinematics indicate  
that the isodensity contours of the total mass distribution 
at $\sim$13~kpc (assuming pure Hubble flow at 
75~km~s$^{-1}$~Mpc$^{-1}$) are consistent with perfect axisymmetry, 
with $(b/a)_\rho = 0.96 \pm 0.08$ (Franx, van~Gorkom \& de~Zeeuw 1994).
The technique applied to the ring IC~2006 can be extended to the 
filled gas disks typical of spirals 
(Schoenmakers, Franx \& de~Zeeuw 1997) and used to study two-dimensional 
HI velocity fields measured via 21cm synthesis mapping.  
The measured value depends on the (unknown) viewing angle with respect 
to the short axis of the potential; averaging over random viewing 
angles yields a mean value of $(b/a)_\rho = 0.85 \pm 0.04$ 
for a sample of 7 field spirals (Schoenmakers 1998, 1999).  

All of this evidence is consistent with $(b/a)_\rho \gtorder 0.7$ 
at the edge of the optical disk or just beyond.  
Random viewing angles would cause the gas in oval orbits to be viewed 
at different positions, inducing scatter in the measured HI linewidth 
at a given absolute magnitude; the small scatter in the Tully-Fisher relation  
limits this ovalness in the total density to 
$(b/a)_\rho > 0.7$ (Franx \& de~Zeeuw 1994). 
Given that there are other sources for scatter in the Tully-Fisher 
relation, the true constraint may lie closer to $(b/a)_\rho > 0.84$. 
Similarly, given other possible sources of non-axisymmetry, such 
as warping, possible non-equilibrium gas orbits (due to recent accretion), 
and bar and spiral structure, most values in Table~\ref{bona.table} 
should be considered lower limits to the axis ratio of the dark density 
distribution, which is more likely to be $(b/a)_\rho \gtorder 0.8$ 
near the outskirts of spirals. 

\section{Flattening of Dark Halos: $(c/a)_\rho$}

The question of whether dark matter is distribed in a spherical 
$(c/a)_\rho = 1$ halo, a flattened halo, or a  
disk-like $(c/a)_\rho \ltorder 0.2$ structure is important not only 
because of the implications for galactic dynamics, but also because of the 
suggestion that galactic dark may be baryonic and dissipative, 
e.g., clumps of cold gas in the disk (Pfenniger, Combes \& Martinet 1994). 
The question cannot be answered by traditional rotation curve analysis; 
luminous tracers capable of probing of the vertical gradient of the 
gravitational potential are required. 
In Table~\ref{cona.table} several such tracers that have been used 
for this purpose are listed together with the vertical-to-equatorial 
axis ratio $(c/a)_\rho$ inferred from their study.

The anisotropy in the velocity dispersion of Population II halos stars 
has been used by several authors to estimate the flattening of the 
Milky Way mass distribution near the solar radius 
(Binney, May \& Ostriker 1987, van~der~Marel 1991, Amendt \& Cuddleford 1994).  
Due to uncertainties in Galactic parameters and particularly the 
orbital structure of the tracer stars, these studies have remained 
largely inconclusive, with results in the range $0.3 < (c/a)_\rho < 1$. 
Modeling the velocity distribution of local stars with Hipparcos proper 
motions has resulted in the constraint that the local scale height of the 
total mass must be larger than $\sim 2$kpc, while the density scale 
length is estimated to be $1.8 \pm 0.2$kpc (Bienaym\'e 1999).  
These measurements are complicated by indications that the flattening of
the tracer population and the shape of its orbit structure may change 
with radius (e.g., Sommer-Larsen \etal\ 1997, and references therein.)

\begin{table}
\caption{Observational Constraints on Dark Halo Flattening} \label{cona.table}
\begin{center}\scriptsize
\begin{tabular}{lcrr}
\multicolumn{1}{l}{Method/Tracer} & \multicolumn{1}{c}{System}  
& \multicolumn{1}{c}{z Extent Probed} & \multicolumn{1}{c}{($c/a$)$_\rho$} \\
~~~~Reference & & \multicolumn{1}{c}{(kpc)} & \\
\tableline
 & & & \\
Population II Kinematics & & & \\
~~~~Binney, May \& Ostriker 1987 & Milky Way & $\sim$20~kpc & $0.3 - 0.6$ \\
~~~~van~der~Marel 1991 & Milky Way & $\sim$20~kpc & $> 0.34$ \\
~~~~Amendt \& Cuddleford 1994 & Milky Way & $<$50~kpc & $\sim 0.7$ \\
3-D Kinematics of Local Disk Stars & & & \\
~~~~Bienaym\'e, O. 1999 & Milky Way & --- 
	& $(z_0)_\rho > 2$~kpc\tablenotemark{a}  \\
Geometry of X-ray Isophotes & & & \\
~~~~Buote \& Canizares 1996 & S0 NGC 1332 & $\sim 15$~kpc & $0.28 - 0.53$ \\
~~~~Buote \& Canizares 1997 & E4 NGC~720 & $\sim 10$~kpc & $0.37 - 0.60$ \\
~~~~Buote \& Canizares 1998 & E4 NGC~3923 & $\sim 13$~kpc & $0.34 - 0.65$ \\
Kinematics of Polar Rings & & & \\
~~~~Arnaboldi \etal\ 1993 & E4 AM2020-504 & $\sim 5$~kpc & $\sim 0.6$ \\
~~~~Sackett \etal\ 1994 & E7 NGC~4650A & $\sim 17$~kpc & $0.3 - 0.4$ \\
~~~~Sackett \& Pogge 1995 & E4/5 A0136-0801 & $\sim 12$~kpc & $\sim 0.5$ \\
~~~~Combes \& Arnaboldi 1996 & E7 NGC~4650A & $\sim 17$~kpc & $> 0.2$ \\
Precession of accreted dusty gas disk & & & \\
~~~~Steiman-Cameron, Kormendy \& Dusisen & SO NGC~4753 & few kpc & $0.84 - 1.0$\\
Evolution of Gaseous Warps & & & \\
~~~~Hofner \& Sparke 1994 & 5 Spirals  & few kpc & $0.6 - 0.9$ \\
~~~~New \etal\ 1998 & 5 Spirals & few kpc & $> 0.85$\\
Flaring of HI Gas Layer of Spirals & & & \\
~~~~Olling 1996 & Sc NGC~4244 & few kpc & $0.1 - 0.5$ \\
~~~~Olling \& Merrifield 1997 & Milky Way & few kpc & $0.5 - 1.0$ \\
~~~~Becquaert \& Combes 1997 & NGC~891 & few kpc & $0.2 - 0.5$ \\
Strong Radio Lensing of Spirals & & & \\
~~~~Koopmans, de Bruyn \& Jackson 1998 & B1600+434 & --- & $> 0.4$ \\
\end{tabular}
\end{center}
\vglue -0.7cm
\tablenotetext{a}{Here $(z_0)_\rho$ is the disk scale height of the total density.}
\end{table}

Buote \& Canizares (1996, 1997, 1998) have measured 
the flattening of extended X-ray isophotes which, if the gas 
is in hydrostatic equilibrium, should trace the 
shape of the underlying potential in the early-type hosts. 
Their results, which are independent of assumptions about X-ray gas 
temperature, indicate substantially flattened dark mass 
axis ratios of $0.28 < (c/a)_\rho < 0.65$, equal 
to or more flattened than the optical light in each system.
Isophote twists in NGC~720 indicate a 
possible misalignment of the luminous and dark components.

Accreted gas appears to settle onto orbits that are near the equatorial 
or polar planes of galaxies, suggesting that the underlying potential 
is not spherical.  The shape and kinematics of polar rings are  
excellent probes of $(c/a)_\rho$;  
these rare rings of gas and stars can extend to 20 stellar disk 
scale lengths in orbits nearly perpendicular to the 
equatorial stellar plane.  Pioneering work on the kinematics of 
polar rings yielded slightly flattened halos with very large uncertainties 
(Schweizer, Whitmore \& Rubin 1983, Whitmore, McElroy \& Schweizer 1987). 
Subsequent studies using more detailed mass models 
and higher quality data over a larger radial 
range narrowed the range of dark density flattenings to 
$0.3 \ltorder (c/a)_\rho \ltorder 0.6$ (Arnaboldi \etal\ 1993, Sackett 
\etal\ 1994, Sackett \& Pogge 1995).  
In each case, the inferred dark halo flattening is aligned with and 
comparable to that of the stellar body, 
as illustrated for NGC~4650A in Fig.~1.

\begin{figure}
\vglue -3.5cm
\hglue -4.25cm
\epsfysize=18cm\epsffile{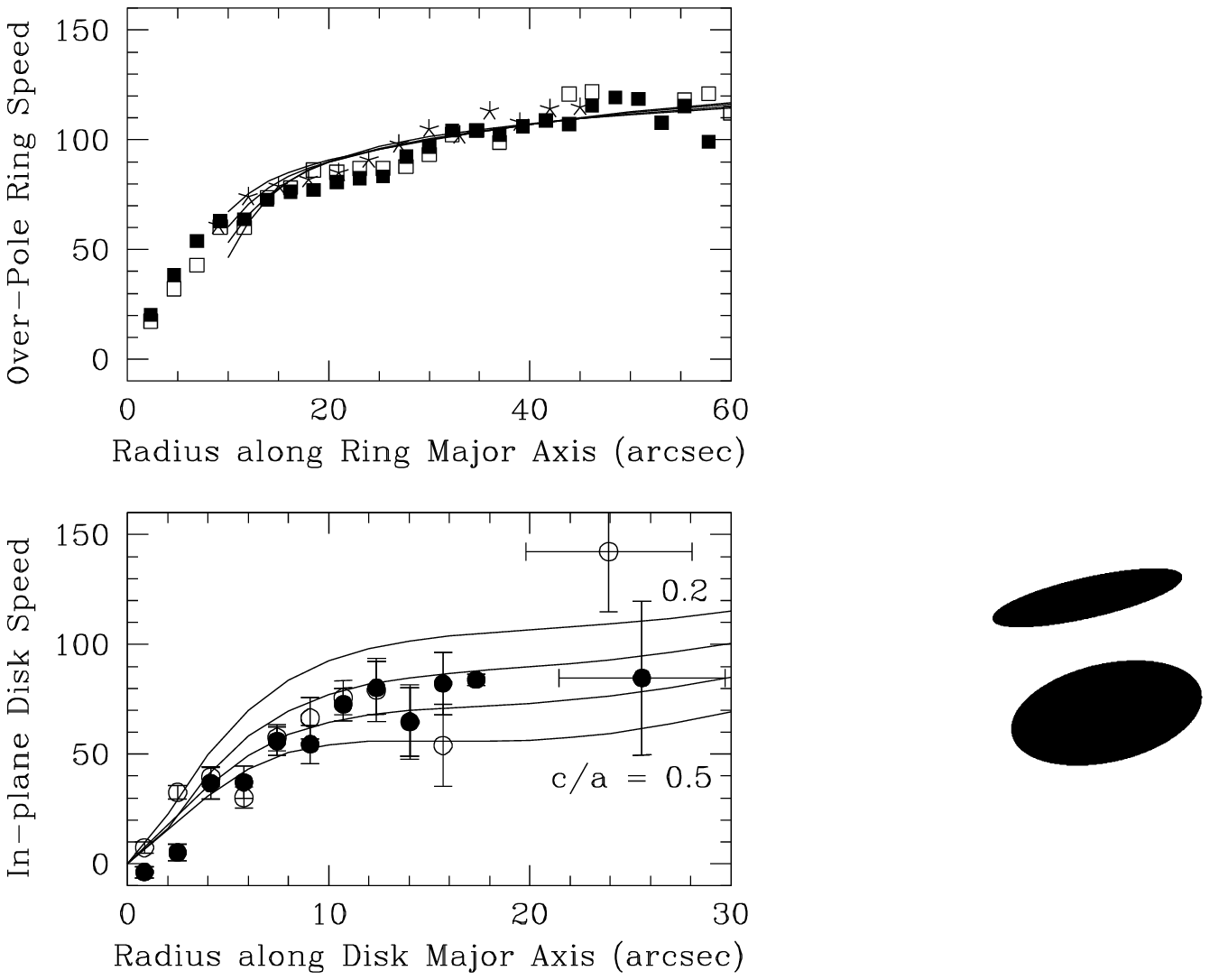}
\vglue -14.75cm
\hglue 8.5cm
\epsfysize=5cm\epsffile{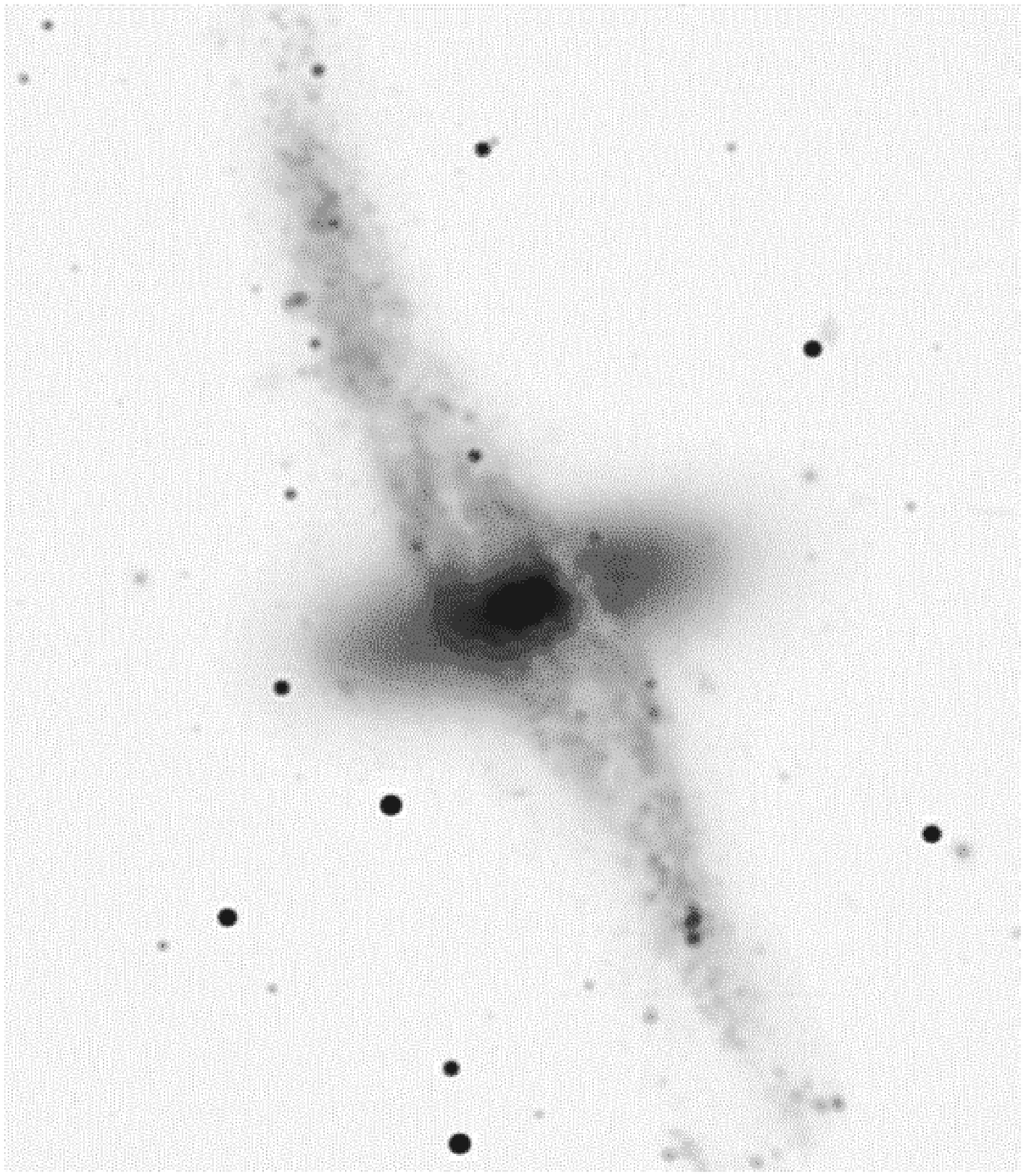}
\vglue 4.75cm
\caption{
{\it Left:\/} Mass models for polar ring galaxy NGC~4650A 
with different dark halo flattening are 
adjusted to fit the ring kinematics (top).  Only those 
with $0.3 < c/a < 0.4$ are able to match disk star kinematics of the 
main galaxy (bottom). 
{\it Top Right:\/} NGC~4650A as seen in one 
of the first images of the VLT at the European Southern Observatory.  
{\it Bottom Right:\/}  Morphology of dark halo shapes considered in left 
panel; the best-fitting halo shape is intermediate to those shown.
(Adapted from Sackett \etal\ 1994).
}\label{n4650a.fig}
\end{figure}

Combes \& Arnaboldi (1996) have also suggested that the dark matter 
distribution of NGC~4650A may be highly flattened, but toward the 
plane of the ring itself and thus 
misaligned from the flattened stellar body by 90$\deg$.  Their ring 
model requires a stellar mass-to-light ratio of $M/L_B = 5$ and 
a constant inclination angle of 85$\deg$ 
(to produce a large line-of-sight correction to the ring velocities), 
that appear to be inconsistent with the color of the 
ring and its inclination as derived from 
optical imaging and H$\alpha$ velocity fields.  

The substantially flattened halos inferred from X-ray gas and polar 
ring studies contrast sharply with those inferred from the kinematics 
of gas disks close to the equatorial plane.  In order to fit the 
complicated pattern of dust lanes in the SO NGC~4753, a precessing disk 
model was used by Steiman-Cameron, Kormendy \& Durisen (1992). 
The flattening is coupled to the assumed viscosity and 
age of the structure: if it has 
completed 6 or more orbits at all radii, the flattening must be 
quite modest, $(c/a)_\rho \geq 0.84$ to avoid excessive winding. 

New \etal\ (1998) have applied a similar precessing, viscous, disk model  
in scale-free potentials to fit the warps observed in five 
spirals.  They find that $\eta_\Phi$, a measure of the strength of 
the quadrupole moment, must be small ($10^{-3} < \eta_\Phi < 10^{-2})$,  
implying density axis ratios of 
$0.85 < (c/a)_\rho \ltorder 1 $  
for nearly axisymmetric structures older than one-tenth a Hubble time.  
Warp models that begin with a preformed gas disk that evolves gravitationally 
toward a discrete bending mode in a tilted rigid dark halo 
can also reproduce the observed warping of five spirals --- as long as  
the dark halo is not too flattened, ie., 
$0.6 \ltorder (c/a)_\rho \ltorder 0.9$ (Hofner \& Sparke 1994). 
The longevity of misaligned warps in tilted halos may be 
limited, however; simulations with responsive halos indicate 
that the disk and inner halo find a common plane in a few orbital 
times (Dubinski \& Kuijken 1995).  Furthermore, Debattista \& Sellwood (1999) 
argue that misalignment of gas and halo angular momentum 
can create transient warps with appropriate characteristics even 
in spherical halos.  

The thickness of the HI gas layer in spirals is due (at least in part) to 
the hydrostatic balance between random gas motion and the 
vertical restoring force due to the total gravitational potential.
By modeling the distribution and velocity of gas in the edge-on 
Sc NGC~4244, and taking into account systematic effects on the 
measured flaring due to inclination and extragalactic ionizing radiation, 
an axis ratio $(c/a)_\rho = 0.2^{+0.3}_{-0.1}$  
is derived for the dark density in the region of the flare (Olling 1996). 
A portion of the uncertainty is due to the unknown  
anisotropy of the gas motions.
A similar analysis for the Milky Way yields a considerably 
rounder halo $(c/a)_\rho = 0.7 \pm 0.3$ (Olling \& Merrifield 1997).  
Applied to the edge-on NGC~891, the method gives substantially flattened 
density axis ratios of  $0.2 < (c/a)_\rho < 0.5$ (Becquaert \& Combes 1998), 
although non-flare models for the HI distribution and kinematics 
have also been put forward for this galaxy 
(Swaters, Sancisi \& van~der~Hulst 1997). 

\begin{figure}
\hglue 0cm
\epsfysize=6cm\epsffile{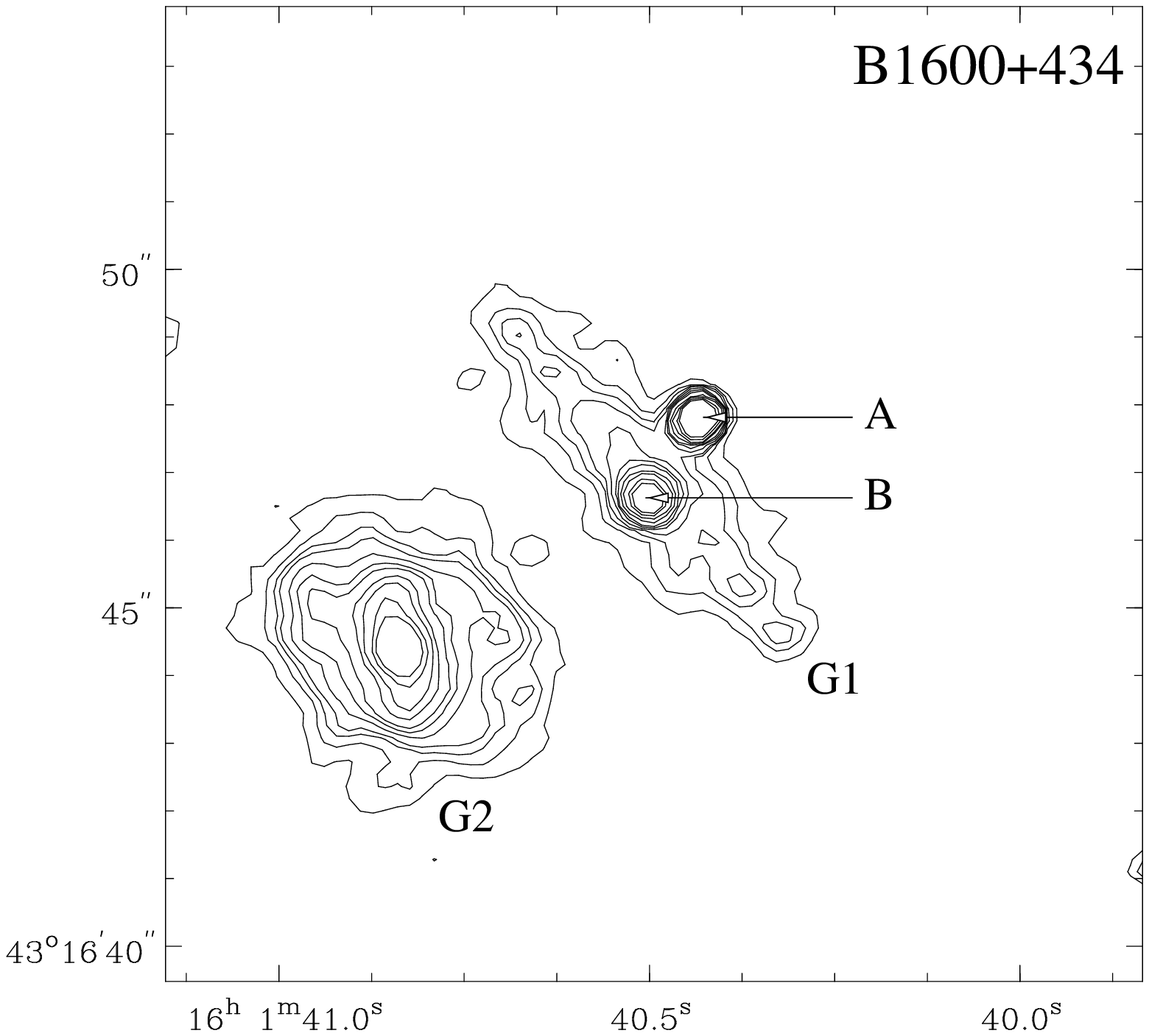}
\vglue -6.25cm
\hglue 6.75cm
\epsfysize=6.5cm\epsffile{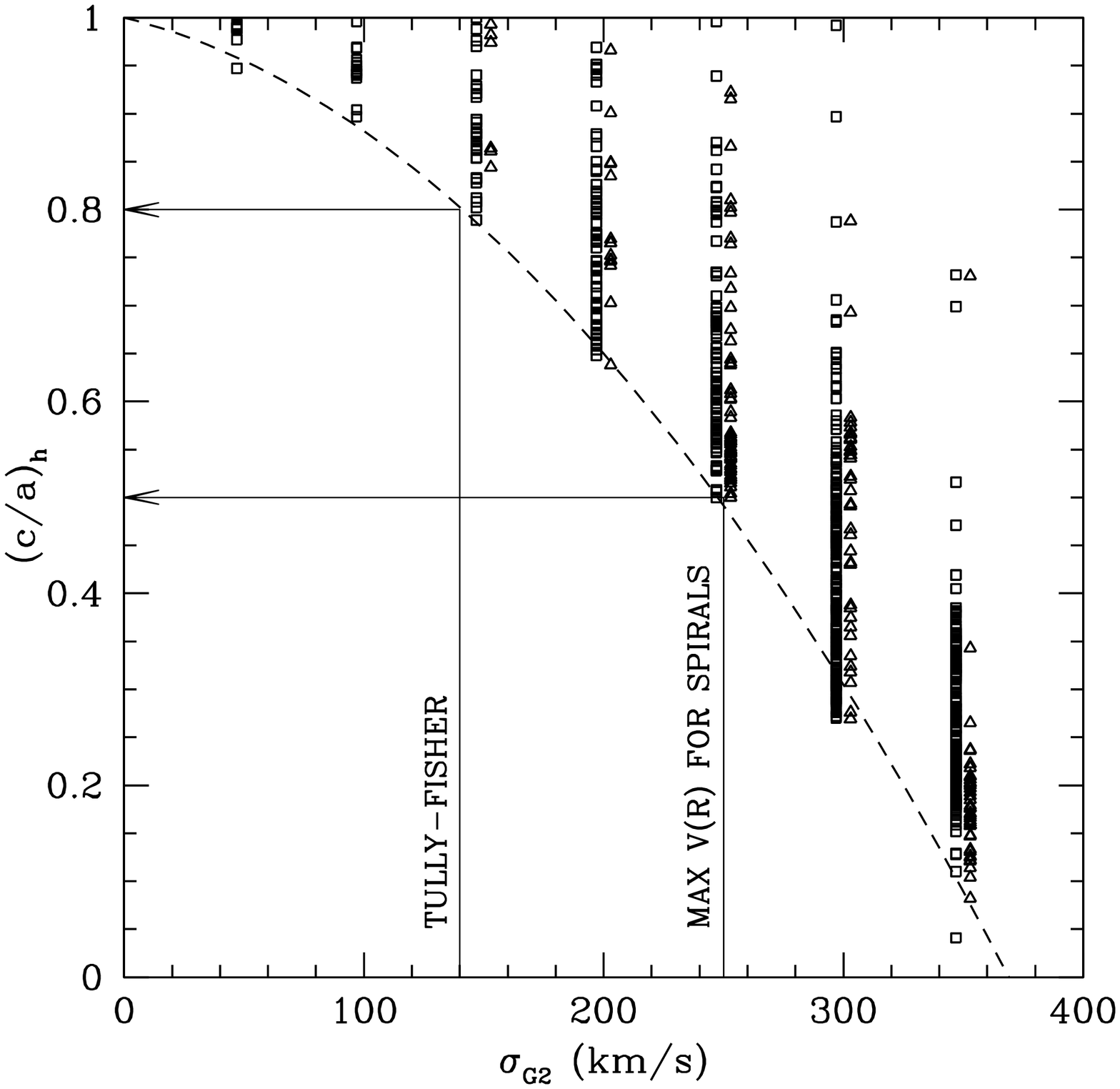}
\caption{
{\it Left:\/} Deconvolved Nordic Optical Telescope R-band image of the CLASS
gravitational lens B1600+434. The edge-on lens galaxy (G1) is seen
between the images A and B of the background radio source. 
The companion galaxy (G2) is located 4.5 arcsec southeast of G1. 
{\it Right:\/}
The flattening (c/a)$_h$ of G1 dark halo is plotted against
the velocity dispersion of the companion galaxy G2. Squares and
triangles indicate a range of halo and galaxy (G1 and G2) models that 
fit the data. (Adapted from Koopmans, de~Bruyn \& Jackson 1998.)
}\label{leon.fig}
\end{figure}

Strong gravitational lensing provides a new way 
to measure the flattening of dark halos since the image positions and relative 
brightnesses are altered by the ellipticity of the foreground 
lensing potential.  
Lensing is sensitive to the total mass column 
along the line of sight, so modeling 
of luminous mass of the galaxy and any nearby dark masses 
(e.g., external shear) is required in order to derive accurate results, but   
the equilibrium assumptions required in most dynamical mass estimators 
play no role.  Lens models often require substantially flattened 
density distributions with 
$(c/a)_\rho \ltorder 0.4$ for strong galactic lenses 
(Kochanek 1995, and references therein).  The magnitude of the shear 
term required to reduce ellipticities to those representative 
of E and S0 galaxies may require internal shears due to 
misaligned dark and luminous lens mass (Keeton, Kochanek \& Seljak 1997).  
In some cases, strict limits 
can be placed on the flattening of the dark lensing density, even in 
the presence of known external shear, as indicated in Fig.~2 which   
displays modeling results for a late-type lensing spiral at redshift z=0.415 
(Koopmans, de~Bruyn \& Jackson 1998).
Lower limits on the dark halo flattening were obtained by 
constraining the velocity dispersion (and thus external shear) 
of the neighboring galaxy with the Tully-Fisher relation and, more conservatively,
with the maximum rotation velocity of massive spirals.

\section{Summary and Discussion}

The dark halo shapes discussed here were inferred in different 
types of galaxies, on different spatial scales, with  
methods that have different systematic 
uncertainties and assumptions.  Some systems may present special advantages 
that allow techniques not possible elsewhere, but may not be 
representative of the general population.  
Measurements at smaller radii may have the luxury 
of a wider variety of tracers and higher quality data, 
but are contaminated by the influence of the luminous mass. 
Inferred shape parameters derived at large distances  
safely can be assumed to probe the dark matter distribution, but may  
assume an equilibrium condition that is less likely to satisfied 
at the large galactocentric radius of the tracer.

The contribution of the dark matter to the inner rotational support of 
galaxies is still not fully understood, although it appears that it 
is dominant in fainter, lower surface brightness galaxies.  
Several lines of evidence suggest that halos extend to $50 - 100$~kpc; 
whether they extend to 200~kpc and beyond involves assessing 
conflicting evidence and model-dependent conclusions. 
  
The observational evidence taken together indicates that 
total density distribution near the optical radius ($\sim 20$~kpc) 
in spirals has an in-plane axis ratio $(b/a)_\rho \gtorder 0.8$.  
If any of the measured ovalness is due to sources 
other than dark halo shape, dark matter halos at this radius may 
be even more axisymmetric.  At least near the stellar body itself, 
dark halos are not strongly triaxial.  

The inferred values for halo flattening are more disparate.   
Techniques that measure $(c/a)_\rho$ rather close to the equatorial 
plane of the stellar body return both quite spherical (precessing warps) 
and substantially flattened (gas layer flaring) results, even though 
they are applied to systems that appear to be quite similar. 
One might imagine that the dark potential is preferentially flattened 
near the equatorial plane due to the influence of the massive stellar 
disk; the spherical results returned by the warp analysis therefore  
seem to indicate that effect of the disk is negligible.    
Perhaps other mechanisms, like continual 
accretion, are responsible for shaping the distribution and 
velocity structure of the outer neutral hydrogen layer.   
It does seem clear that the gradient of the mass distribution near the 
galactic plane is not steep enough to suggest that the dark mass 
lies in a thin ($<$2~kpc) disk. 

At heights $10 -20$~kpc above the equatorial plane, Milky Way halo 
star kinematics, the geometry of X-ray isophotes, and polar ring 
studies all indicate that the dark mass is substantially flattened 
with $0.3 < (c/a)_\rho < 0.7$.  At these distances, 
there is some indication that for early-type systems the 
flattening of the dark and luminous mass may be similar.  Polar rings 
with flattened halos may be more likely to capture material onto 
polar orbits, but this selection effect is strongly countered by the 
short lifetime expected for differentially precessing rings accreted 
at a slightly skew angle in a non-spherical potential.  

How do theoretical expectations square with these observational facts?
N-body simulations of dissipationless collapse produce strongly 
triaxial dark halos (Frenk \etal\ 1988, Dubinski \& Carlberg 1991, 
Warren \etal\ 1992) that are both oblate and prolate and thus at strong 
odds with the data.  Adding a small fraction ($\sim$10\%) of dissipative 
gas, however, results in halos of a more consistent shape --- 
nearly oblate, $(b/a)_\rho \gtorder 0.8$, 
but moderately flattened, $(c/a)_\rho = 0.5 \pm 0.15$ 
(Katz \& Gunn 1991, Dubinski 1994, Dubinski \& Kuijken 1995).
The dark halo shapes inferred from observations are thus consistent with 
expectations from cold dark matter simulations, as long as 
dissipative baryons form at least $\sim$10\% of the total mass 
within the volume relevant for the formation of the stellar galaxy.

\section{A Look to the Future}

What can we expect as we look toward the future?  Doubtless many of 
the techniques discussed here will be improved, and new ones invented. 

Harmonic analysis of two-dimensional velocity fields is likely 
to provide more detailed information about the non-axisymmetry of 
galactic mass distributions in different environments as 
the effect of spirals arms and warping is quantified.  
Lensing may contribute to the question of dark halo shape at many 
levels because, although hampered with its own difficulties, it is 
insensitive to the (often unknown and/or complicated) 
dynamical state of the system.  Strong lensing, especially 
when applied to radio sources where extinction does not play a role 
in selection or altering image brightness, will become increasingly 
important in constraining the flattening of galactic mass distributions 
(see Fig.~2).  

If a substantial portion of galactic dark halos is 
composed of compact massive objects, microlensing may help determine 
not only the shape but the constitution of dark halos.  
In the near future, the promise of using multiple lines of sight 
(Galactic Bulge, LMC, SMC) to probe the shape of the Milky Way 
microlensing halo (Sackett \& Gould 1993) will be hampered by small 
number statistics and ultimately perhaps by the possibility that a 
substantial number of the microlenses do not lie in the halo.
If a considerable fraction of galactic dark matter {\it is\/} comprised 
of microlenses, projects to map the microlensing optical depth across the 
face of M31 will measure its intrinsic shape (Crotts 1992, Baillon \etal\ 1993).

\begin{figure}
\epsfysize=12.5cm\epsffile{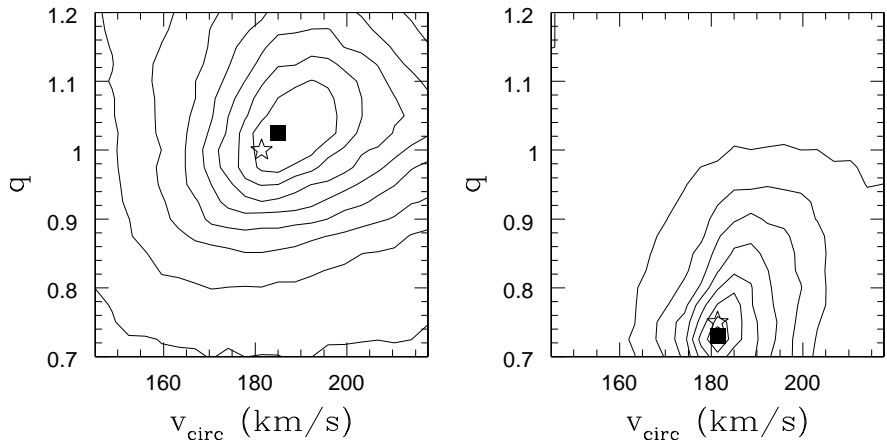}
\hglue 1.25cm
\vglue -6.5cm
\caption{Contours indicate the percentage of particles (90\%, 80\%, 70\% . . .) 
correctly modeled by applying an analytic description of tidal streamers 
in potentials of two different depths and flattenings to simulated SIM data 
supplemented with radial velocities.  The ``true'' circular speed 
and flattening $q = (c/a)_\Phi$ of the potential is indicated with the stars;  
filled squares indicate the recovered halo parameters.  
(Adapted from Johnston \etal\ 1998.)
}\label{kathryn.fig}
\end{figure}

Statistical weak lensing combined with redshift information is likely 
to become more important in constraining the outer radial 
profile of dark halos in a manner that is less model dependent than 
methods that rely on satellite dynamics.  The cold kinematics and 
confined spatial distribution of very extended ($ > 50$~kpc) HI features, 
such as tidal tails and ``Leo rings'' will be a welcome addition 
in assessing the extent of galactic halos, if they can be found in 
sufficient number and characterized. 

Finally, the shape and extent of stellar debris trails from infalling 
satellites may also be an increasingly important probe of halo size 
and shape, especially if it can be combined with kinematical information, 
as may be possible with the launch of the Space Interferometry Mission (SIM) 
to measure proper motions of Galactic stars brighter than 20th magnitude.
Simulations indicate that, if supported by ground-based programs to 
determine radial velocities, SIM-measured proper motions of 100 stars 
in a tidal stream would be sufficient to determine the circular 
velocity and flattening of the Galactic potential to within $\sim 2 - 3$\% 
(Fig.~3) at the orbital radius of the streamer (Johnston \etal\ 1998.)
 
\acknowledgments

I am grateful to the organizing committee for travel support and to 
Leon Koopmans and Kathryn Johnston for preparing figures from their 
work for inclusion in this review.

\end{document}